\documentclass[aps,preprint,floats]{revtex4}
\usepackage{epsf}
\usepackage{graphicx}

\begin{document}

\preprint{hep-ph/0304215}

\title{$A^b_{FB}$ and $R_b$ at LEP and New Right-Handed Gauge Bosons}


\author{Xiao-Gang He}
\email[]{hexg@phys.ntu.edu.tw}
\affiliation{Department of Physics\\National Taiwan University\\
Taipei, Taiwan 10764, R.O.C.}

\author{G. Valencia}
\email[]{valencia@iastate.edu}
\affiliation{Department of Physics\\
Iowa State University\\Ames, IA 50011}

\date{\today}

\begin{abstract}

We explore models with additional right-handed gauge bosons that couple 
predominantly to the third generation in the context of 
$b\bar{b}$ production at LEP. In particular we investigate potential new 
contributions to $\delta g_{Rb}$ that are needed if the measured 
$A^b_{FB}$ at the $Z$ peak is interpreted as a signal of new physics. 
We identify two sources of large $\delta g_{Rb}$ corrections: 
$Z-Z^\prime$ mixing at tree-level, and one-loop effects from a 
new $SU_R(2)$ triplet of gauge bosons. We find that the 
latter can contribute to $\delta g_{Rb}$ at the $1\%$-level. 
We place bounds on the mass of the additional $Z^\prime$ gauge boson 
that occurs in these models using the $R_b$ measurements from LEP-II. 
We find that even in cases where the $Z^\prime$ couples almost exclusively 
to the $b$ and $t$-quarks, masses lighter than about $500$~GeV are already 
excluded. 

\end{abstract}

\pacs{PACS numbers: 12.15.-y, 12.60.-i, 14.70.-e}
\keywords{}

\maketitle


\section{Introduction}

The precision measurements at the $Z$ resonance continue to exhibit 
a deviation from the standard model in the observable $A^{b}_{FB}$ 
by about $-3.2$ standard deviations \cite{Abbaneo:2001ix,Hagiwara:fs}. 
At the same time $R_b$ deviates from the standard model by only $1.4$ 
standard deviations \cite{Hagiwara:fs}. 
It has been pointed out by Chanowitz \cite{Chanowitz:2001bv,Chanowitz:2002cd} 
that the deviation in  $A^{b}_{FB}$ presents a problem for the standard model 
whether it is genuine or not. In particular, Chanowitz argues 
that if the anomaly in $A^{b}_{FB}$ is attributed to systematic error 
and dropped from the LEP fits, then the indirect determination of the 
Higgs mass is in conflict with the direct limit 
\cite{Chanowitz:2001bv,Chanowitz:2002cd}. One possible interpretation of 
this result is that there is new physics associated with the $Zb\bar{b}$ 
couplings, and we explore this possibility in the context of non-universal 
right-handed interactions.

We adopt the following notation for the effective couplings between 
the $Z$-boson and the $b$ and $t$ quarks, 
\begin{eqnarray}
{\cal L} &=& -{g\over 2 \cos\theta_W} \bar f \gamma^\mu 
\left[\left(g_{Lf}+\delta g_{Lf}\right)P_L+
\left(g_{Rf}+\delta g_{Rf}\right)P_R\right]
f Z_\mu,
\label{treevertex}
\end{eqnarray}
with $P_{L,R} = (1\mp\gamma_5)/2$, and tree-level standard model couplings 
$g_{Lf}$ and $g_{Rf}$ as in the Appendix. In terms of these effective 
couplings, the results in Ref.\cite{Abbaneo:2001ix} suggest that new 
physics could be responsible for as much as $\delta g_{Rb} \sim 0.04$, 
$\delta g_{Lb} \sim 0.004$ . 
At the same time, new physics contributions to the $\tau$-lepton 
couplings are constrained to be at most at the $0.001$ level.

Several discussions of new physics effects regarding $A^{b}_{FB}$ 
or $R_b$ have appeared in the literature. Among them: light SUSY partners 
\cite{Altarelli:2001wx}; quark mixing with new fermions  
\cite{Bamert:1996px,Chang:1996pf,Chang:1998uj,Choudhury:2001hs}; 
top-color \cite{Burdman:1997pf,Yue:2000ay,Triantaphyllou:1998ke}; 
top-flavor \cite{Muller:1996qs,Malkawi:1999sa,Malkawi:1996fs,Lynch:2000md}; 
and non-universal left-right models \cite{He:2002kn}. Our goal in this paper 
is to extend our results in Ref.~\cite{He:2002kn} by computing the 
dominant one-loop effects to $\delta g_{Rb}$, and by using the LEP-II 
data on $e^+e^- \rightarrow b\bar{b}$ to constrain the mass of the new 
gauge bosons.

In the standard model the one-loop corrections to $\delta g_{Lb}$ 
that are proportional to $M_t^2$ are approximately $0.006$
\cite{Altarelli:hv}. We can use this result as a benchmark for 
$\delta g_{Rb}$ from new physics, suggesting that if it is to 
occur at one-loop there must be an enhancement relative to 
the standard model electroweak corrections. This is precisely what can 
occur in models such as those we discussed in Ref.~\cite{He:2002kn}, where 
the coupling strength of the new right-handed interaction, $g_R$, is 
significantly larger than the $SU(2)_L$ coupling $g_L$. In this paper 
we calculate these corrections in a simple case and find that 
$\delta g_{Rb}$ from one-loop effects can be $1\%$.

Our paper is organized as follows. In Section II we describe models 
with additional right-handed gauge bosons that could change the value 
$\delta g_{Rb}$ significantly while respecting other phenomenological 
constraints. In particular we discuss several ways in which the 
predominant effects occur for the $b$ and $t$-quark couplings but not for 
the $\tau$-lepton couplings through $Z-Z^\prime$ mixing. 
In Section III we show how, even in the absence of $Z-Z^\prime$ mixing, 
one-loop contributions to $\delta g_{Rb}$ can occur at the $1\%$-level. 
In Section IV we present bounds on the mass of the new gauge bosons 
from the process $e^+ e^- \rightarrow b\bar{b}$ at LEP-II. 
We state our conclusions in Section V and relegate some details to the Appendix.

\section{Non-Universal Left-Right Models} \label{secmods}

The models to be discussed are variations of  
left-right models \cite{lrmodel,babu} in which the 
right-handed interactions single out the third generation. Our basic 
model was introduced in Ref.~\cite{He:2002kn} and we start by recalling 
its salient features. 
The gauge group of the model is $SU(3)\times SU(2)_L\times SU(2)_R
\times U(1)_{B-L}$ with gauge 
couplings $g_3$, $g_L$, $g_R$ and $g$, respectively. 
The model differs from other left-right models in 
the transformation properties of the fermions. 

The first two generations are chosen to have the same 
transformation properties as in the standard model with $U(1)_Y$ replaced 
by $U(1)_{B-L}$,
\begin{eqnarray}
&&Q_L = (3,2,1)(1/3),\;\;\;\;U_R = (3,1,1)(4/3),\;\;\;\;D_R = (3,1,1)(-2/3),
\nonumber\\
&&L_L = (1,2,1)(-1),\;\;\;\;E_R = (1,1,1)(-2).
\label{gens12}
\end{eqnarray}

The numbers in the first parenthesis are the $SU(3)$, $SU(2)_L$
and $SU(2)_R$ group representations respectively, and the number in the second 
parenthesis is the $U(1)_{B-L}$ charge.

The third generation is chosen to transform differently,
\begin{eqnarray}
&&Q_L(3) = (3,2,1)(1/3),\;\;\;\;Q_R(3) = (3,1,2)(1/3),\nonumber\\
&&L_L(3) = (1,2,1)(-1),\;\;\;\;L_R = (1,1,2)(-1).
\label{gen3}
\end{eqnarray}

The above assignments are unusual compared with the conventional left-right
model, but they enhance the difference 
between the right handed couplings of the first two and the third generations.
This model is anomaly free.

The correct symmetry breaking and mass generation of particles 
can be induced by the vacuum expectation values of
three Higgs representations: 
$H_R = (1,1,2)(-1)$, whose non-zero vacuum expectation value (vev) $v_R$
breaks the group down to $SU(3)\times SU(2)\times
U(1)$; and the two Higgs multiplets, 
$H_L = (1,2,1)(-1)$ and $\phi = (1,2,2)(0)$, which break the symmetry 
to $SU(3)\times U(1)_{em}$. For the purpose of symmetry breaking, only 
one of $H_L$ or $\phi$ is sufficient, but both are required 
to give masses to all fermions. It is possible to introduce additional 
Higgs representations as mentioned in Ref.~\cite{He:2002kn}, but we 
will not do so in this paper.

The introduction of $\phi$ causes the standard model $W_L$ and $Z_o$ to mix 
with the new $W_R$ and $Z_R$ gauge bosons. Here $W_R$ is the $SU(2)_R$
charged gauge boson and $Z_R$ is a linear combination of
the neutral component of the $SU(2)_R$ gauge boson $W_{3R}$ and
the $U(1)_{B-L}$ gauge boson $B$. Specifically, 
\begin{eqnarray}
&&Z_o = \cos\theta_W W_{3L} - \sin\theta_W \cos\theta_R B - 
\sin\theta_W \sin\theta_R W_{3R},\nonumber\\
&&Z_R = \cos\theta_R W_{3R} - \sin\theta_R B,
\end{eqnarray}
where $\tan \theta_R = g/g_R$.

In the bases $(W_L,\;W_R)$ and $(Z_o,\;Z_R)$ for the massive
gauge bosons, the mass matrices were given in Ref.~\cite{He:2002kn} 
and we reproduce them here for later convenience:
\begin{eqnarray}
m^2_{11W} &=& 
{1\over 2}g^2_L (|v_L|^2 + |v_1|^2 + |v_2|^2),
\, \, 
m^2_{22W} \, = \, {1\over 2} g_R^2
(|v_R|^2 + |v_1|^2 + |v_2|^2),
\nonumber\\
m^2_{12W} &=& -g_Lg_R Re(v_1v_2^*), \, \,
m^2_{11Z} \, = \, 
{1\over 2}{g^2_L\over \cos^2\theta_W} 
(|v_L|^2  + |v_1|^2 + |v_2|^2),
\nonumber\\
m^2_{22Z} &=& {1\over 2} {g^2_R \over \cos^2\theta_R}
(|v_L|^2  \sin^4\theta_R
+ (|v_1|^2 + |v_2|^2) \cos^4\theta_R + |v_R|^2 ),
\nonumber\\
m^2_{12Z} &=& {1\over 4}{g_Lg_R}{\sin\theta_R\over \cos\theta_W}
(  |v_L|^2  \tan\theta_R 
-(|v_1|^2+|v_2|^2)\cot\theta_R )).
\label{massmat}
\end{eqnarray}

After diagonalization of the mass-squared matrices, the lighter and heavier 
mass eigenstates $(Z,\;Z^\prime)$ and $(W,\;W^\prime)$
are given by
\begin{eqnarray}
\left ( \begin{array}{l}
W\\
W^\prime
\end{array}
\right ) 
= \left ( \begin{array}{ll}
\cos\xi_W&\sin\xi_W\\
-\sin\xi_W&\cos\xi_W
\end{array}
\right )
\left (\begin{array}{l}
W_L\\
W_R
\end{array}
\right ),\;\;\;\;
\left ( \begin{array}{l}
Z\\
Z^\prime
\end{array}
\right ) 
= \left ( \begin{array}{ll}
\cos\xi_Z&\sin\xi_Z\\
-\sin\xi_Z&\cos\xi_Z
\end{array}
\right )
\left (\begin{array}{l}
Z_o\\
Z_R
\end{array}
\right ),
\end{eqnarray}
where $\xi_{Z,W}$ are the mixing angles, 
\begin{eqnarray}
\tan(2\xi_{W,Z}) = {2 m^2_{12(W,Z)}\over m^2_{11(Z,W)} - m^2_{22(Z,W)}}.
\end{eqnarray}

In this model there are new interactions between the massive 
gauge bosons and quarks. For the charged current interaction, there are
both left and right handed interactions. In the weak eigenstate basis, 
the charged gauge boson, $W_L$, couples to all generations, but 
the charged gauge boson, $W_R$, only couples to the third generation. 
There is a similar pattern for the neutral gauge interactions. This pattern 
gives rise to interactions between the fermions and the lightest 
physical gauge bosons that can be made to resemble the standard model 
couplings plus enhanced  right-handed couplings for the third generation.  
In the mass eigenstate basis the quark-gauge-boson interactions are 
given by,
\begin{eqnarray}
{\cal L}_W&=& - {g_L\over \sqrt{2}} \bar U_L \gamma^\mu V_{KM} D_L
(\cos\xi_W W^{+}_\mu - \sin\xi_W W^{'+}_\mu)\nonumber\\
&&-{g_R\over \sqrt{2}}
\bar u_{Ri} \gamma^\mu V^{u*}_{Rti}V^{d}_{Rbj} d_{Rj}
(\sin\xi_W W^{+}_\mu + \cos\xi_W W^{'+}_{\mu}) ~+~{\rm h.~c.},
\label{cccoup}
\end{eqnarray}
where $U = (u,\;\;c,\;\;t)$ and $D = (d,\;\;s,\;\;b)$. $V_{KM}$ is
the Kobayashi-Maskawa mixing matrix and $V^{u,d}_{Rij}$ are unitary matrices
which rotate the right handed quarks $u_{Ri}$ and $d_{Ri}$ 
from the weak eigenstate basis 
to the mass eigenstate basis. The repeated indices $i$ and $j$ are summed over
three generations. For the neutral sector the couplings are,
\begin{eqnarray} 
{\cal L}_Z &=& -{g_L\over 2 \cos\theta_W}
\bar q \gamma^\mu (g_V - g_A \gamma_5) q (\cos\xi_Z Z_\mu - \sin\xi_Z Z^\prime_\mu)
\nonumber\\
&+& {g_Y\over 2} \tan\theta_R  
({1\over 3} \bar q_L \gamma^\mu q_L+ {4\over 3} \bar u_{Ri} \gamma^\mu u_{Ri}
-{2\over 3} \bar d_{Ri}\gamma^\mu d_{Ri})
(\sin\xi_Z Z_\mu + \cos\xi_Z Z^\prime_\mu)\nonumber\\
&-& {g_Y\over 2} (\tan\theta_R + \cot\theta_R) (
\bar u_{Ri} \gamma^\mu V^{u*}_{Rti} V^{u}_{Rtj}u_{Rj} - 
\bar d_{Ri} \gamma^\mu V^{d*}_{Rbi} V^{d}_{Rbj} d_{Rj}) 
(\sin\xi_Z Z_\mu + \cos\xi_Z Z^\prime_\mu).
\label{neucoupl}
\end{eqnarray}
In this expression, $g_Y = g \cos\theta_R = g_R \sin\theta_R$,  
$q$ and $q_L$ are summed over  $u,d,c,s,t,b$ quarks, 
and repeated $i,j$ indices are summed over the three generations. 
The first line contains the standard model couplings to the $Z$ 
in the limit $\xi_Z =0$. The first two lines also contain couplings 
of the two $Z$ bosons to quarks that arise through mixing of the 
neutral gauge bosons. 

Similarly, the couplings to leptons are given, in the weak eigenstate basis, by:
\begin{eqnarray}
{\cal L}_Z(lepton)&=& - {g_L\over 2 \cos\theta_W}
\bar \ell \gamma^\mu (g_V - g_A\gamma_5) \ell 
(\cos\xi_Z Z_\mu - \sin\xi_Z Z^\prime_\mu)
\nonumber\\
&+& {g_Y\over 2} \tan\theta_R  
(- \bar \ell_L \gamma^\mu \ell_L -2 \bar E_{Ri} \gamma^\mu E_{Ri})
(\sin\xi_Z Z_\mu + \cos\xi_Z Z^\prime_\mu)\nonumber\\
&-& {g_Y\over 2} (\tan\theta_R + \cot\theta_R) (
\bar \nu_{R \tau} \gamma^\mu \nu_{R \tau} - 
\bar \tau_R \gamma^\mu \tau_R) 
(\sin\xi_Z Z_\mu + \cos\xi_Z Z^\prime_\mu).
\label{coupslep}
\end{eqnarray}
In this case, $\ell$ and $\ell_L$ are summed over $e,\;\mu,\;\tau,\; \nu_e, 
\;\nu_\mu,\; \nu_\tau$ and $E_R$ are summed over three generations.

The most interesting terms in Eqs.~\ref{neucoupl}~and~\ref{coupslep} 
occur in the third line and are 
potentially large if $\cot\theta_R$ is large. 
In the weak interaction basis they affect 
only the third generation whereas in the mass eigenstate basis 
(as written in Eq.~\ref{neucoupl}) they also give rise to flavor changing 
neutral currents. To satisfy the severe constraints that exist on 
flavor changing neutral currents we have to require that the 
$V^d_{R}$ and $V^u_{R}$ matrices be nearly diagonal.

In Ref.~\cite{He:2002kn} we studied the case with $\xi_Z\neq 0$, in 
which $Z-Z^\prime$ mixing is responsible for the shifts in the effective 
right-handed coupling of the $b$-quark. Within this scenario, the model 
given above also induces large shifts in the right-handed coupling of the 
$\tau$-lepton, making it phenomenologically unacceptable. One finds 
for large $\cot\theta_R$\footnote{Notice that there is a factor of 2 
difference between the definitions of $\delta g_{R}$ here and in 
Ref.~\cite{He:2002kn}.},
\begin{eqnarray}
\delta g_{Rb} &\approx & 
-  \sin\theta_W \cot \theta_R V^{d*}_{Rbb}V^d_{Rbb} \, \xi_Z \nonumber \\
\delta g_{Rt} &\approx & 
\sin\theta_W \cot \theta_R V^{u*}_{Rtt}V^u_{Rtt}\, \xi_Z \nonumber \\
\delta g_{R\tau}  &\approx & 
-  \sin\theta_W \cot \theta_R\, \xi_Z.
\label{maindev}
\end{eqnarray}
This last equation constrains the product $\cot \theta_R \xi_Z$ to be 
at the $10^{-3}$ level or less, whereas one would need 
$\cot \theta_R \xi_Z \sim 0.08$  \cite{He:2002kn} 
to explain $A^b_{FB}$ through this 
mechanism. Nevertheless, there are several ways around 
this constraint. One possibility is to eliminate the 
relation between
the $b$-quark and $\tau$-lepton couplings to the new gauge bosons. To 
maintain a model that is anomaly free, this is accomplished by introducing 
additional fermions and can be done in more than one way. Two examples 
are given below. A second possibility is to require the $Z-Z^\prime$ 
mixing to be small (or zero) and in this way satisfy the constraints from 
$\tau$ leptons. As we discuss in Section~\ref{secloop}, 
there is a second mechanism 
at the loop level by which the model can induce significant shifts 
on $\delta g_{Rb}$ and not on $\delta g_{R\tau}$.

We now discuss two ways to modify the model so that it remains anomaly 
free but does not have enhanced couplings for the $\tau$-lepton  in 
the case of large $\cot\theta_R$

\subsection{Modified lepton sector}

In this first example we keep the quark sector as above but make some
modifications to the lepton content. The lepton sector consists of the 
usual three generations (all transforming as in Eq.~\ref{gens12}) plus
\begin{equation}
L_R = \left ( \begin{array}{ll}
\nu'_R\\
e'_R 
\end{array}
\right ) =
(1,1,2)(-1), e'_L = (1,1,1)(-2). 
\end{equation}

Compared with the particle content of Eq.~\ref{gens12}~and~\ref{gen3}, 
the net new particles
are $e'_L$ and $E_R(3)$. Their contributions to gauge anomaly cancel 
each other, and therefore the theory is anomaly free.

The new particle $e'$ can be made heavy because $H_R$ provides its
mass. The neutral new particle $\nu'_R$ can be made heavy by introducing
a $\Delta_R (1,1,3)(-2)$ Higgs representation with large VEV. Therefore, at low
energy one does not need to consider the effect of the new fermions.

The couplings for the usual three generations of leptons become,
\begin{eqnarray}
{\cal L}_Z(lepton)&=& - {g_L\over 2 \cos\theta_W}
\bar \ell \gamma^\mu (g_V - g_A\gamma_5) \ell (\cos\xi_Z Z_\mu - 
\sin\xi_Z Z^\prime_\mu)
\nonumber\\
&+& {g_Y\over 2} \tan\theta_R  
(- \bar \ell_L \gamma^\mu \ell_L -2 \bar E_{Ri} \gamma^\mu E_{Ri})
(\sin\xi_Z Z_\mu + \cos\xi_Z Z^\prime_\mu).
\label{qq}
\end{eqnarray}
Once again $\ell$ and $\ell_L$ are summed over $e,\;\mu,\;\tau,\; \nu_e, 
\;\nu_\mu,\; \nu_\tau$ and $E_R$ is summed over three generations.
The couplings of the new leptons  are
\begin{eqnarray}
{\cal L}_Z(lepton)&=& 
-{g_R\over \sqrt{2}} [\bar \nu'_R \gamma_\mu e'_R 
W_R^\mu + H.c.]\nonumber\\
&-&{g_L\over 2\cos\theta_W}
(-2q \sin^2\theta_W) \bar e' \gamma^\mu e' 
(\cos\xi_Z Z_\mu - \sin\xi_Z Z^\prime_\mu)
\nonumber\\
&+& {g_Y\over 2} \tan\theta_R  
(-2 \bar e' \gamma^\mu e')
(\sin\xi_Z Z_\mu + \cos\xi_Z Z^\prime_\mu)\nonumber\\
&-& {g_Y\over 2} (\tan\theta_R + \cot\theta_R) (
\bar \nu'_{R} \gamma^\mu \nu'_{R } - 
\bar e'_R \gamma^\mu  e'_R) 
(\sin\xi_Z Z_\mu + \cos\xi_Z Z^\prime_\mu).
\end{eqnarray}

\subsection{Modified quark sector}

In this case we have three generations of leptons transforming as in 
Eq.~\ref{gens12} with couplings as in Eq.~\ref{qq}, and we introduce 
additional quarks to cancel the anomalies:
\begin{eqnarray}
Q'_L = \left ( \begin{array}{ll}
u'_L \\
d'_L
\end{array}
\right ) = (3,1,2) (1/3), u'_R = (3,1,1)(4/3), 
d'_R = (3,1,1)(-2/3).
\end{eqnarray}
The usual three generations of quarks have the same quantum numbers 
as in Eqs.~\ref{gens12},~\ref{gen3} and couplings as in 
Eqs.~\ref{cccoup},~\ref{neucoupl}. 
Again the above particle content gives a gauge anomaly free theory, and 
the new particles can be made heavy  because they receive
their mass from the VEV of $H_R$.
The new quarks have couplings,
\begin{eqnarray}
{\cal L}_Z(quark) &=& 
-{g_L\over 2 \cos\theta_W} (-2 q \sin^2\theta_W) \bar q' \gamma^\mu 
q' ( \cos\xi_Z Z_\mu - \sin\xi_Z Z^\prime_\mu)
\nonumber\\
&+& {g_Y\over 2} \tan\theta_R  
({4\over 3} \bar u' \gamma^\mu u'
-{2\over 3} \bar d' \gamma^\mu d')
(\sin\xi_Z Z_\mu + \cos\xi_Z Z^\prime_\mu)\nonumber\\
&-& {g_Y\over 2} (\tan\theta_R + \cot\theta_R) (
\bar u'_{L} \gamma^\mu u'_{L} - 
\bar d'_{L} \gamma^\mu d'_L) 
(\sin\xi_Z Z_\mu + \cos\xi_Z Z^\prime_\mu).
\label{moreqcoup}
\end{eqnarray}

\subsection{Discussion}

The previous two examples illustrate how it is possible to single out 
the $b$ and $t$-quarks with a new right-handed interaction without 
affecting the $\tau$ lepton very much. The price paid is, of course, 
the introduction of additional fermions. The additional fermions can be  
made heavy and this allows us to ignore them at this stage, where 
we are interested only in the effect of potentially strong right-handed 
couplings of the $b$ and $t$ quarks in LEP observables. The new heavy 
fermions are only used to illustrate that it is possible to construct a 
renormalizable, anomaly free, model of this type. 

The couplings of the $b$-quark to the new right-handed gauge bosons remain 
as in the original model so that, according to Eq.\ref{maindev}, 
we require $\xi_Z \cot\theta_R \sim 0.08$ to explain $A_{FB}^b$
\cite{He:2002kn} through $Z-Z^\prime$ mixing. 
With $\cot\theta_R$ large, the new physics effects mainly affect 
the third generation of quarks; our model is in some sense ``leptophobic''.

In Ref.~\cite{He:2002kn}, we pointed out that the process 
$b \rightarrow s \gamma$ severely constrains the mixing of the charged 
gauge bosons $\xi_W$. This constraint is not in conflict 
with the mixing needed in the neutral sector, $\xi_Z$, to fit 
$A_{FB}^b$ as discussed in  Ref.~\cite{He:2002kn}. Here we point out that 
there is another way to obtain Eq.~\ref{maindev} without 
affecting $b \rightarrow s \gamma$. This involves a new model in which the 
$SU(2)_R$ is replaced by a $U(1)_R$ with up quarks (leptons) and down quarks 
(leptons) in $SU(2)_R$ doublets carrying 1 and -1 of $U(1)_R$ charges, 
respectively. This model will also give $\delta g_{Rb} \sim \xi_Z
\cot\theta_R$ as in Eq.~\ref{maindev}, 
but it now arises in the context of models where (a) there are no large 
contributions to $Z \rightarrow \tau^+ \tau^-$, and (b) there are no 
new charged gauge bosons $W^\prime$, so that there are no constraints 
from $b \rightarrow s \gamma$. The contributions to the parameter $T$ 
that occur through mixing of the $Z$ and $Z^\prime$ are identical to 
Ref.~\cite{He:2002kn} and lead to the allowed region of Figure~1 in that 
reference. 

The most important new feature common to all the models that we have 
discussed is the existence of a new $Z^\prime$ gauge boson which 
has enhanced couplings to top and bottom quarks (and perhaps to the 
$\tau$-lepton provided its mixing with the $Z$ is sufficiently
small). In Section~\ref{seclep2} we explore the bounds that exist 
on the mass of this $Z^\prime$ from LEP-II measurements.

\section{One-loop contributions to $\delta g_{Rb}$} \label{secloop}

In models like the ones presented in Section~II, with a new 
$SU(2)_R$ gauge interaction, there is a one-loop 
contribution to $\delta g_{Rb}$ that is present even when there is 
no mixing. A priori we can expect this contribution to be similar 
in size to the standard model contribution to $\delta g_{Lb}$ proportional to 
$M_t^2$. One can imagine a suppression of 
the form $(M_W/M_{W_R})^2$ with respect to the standard model $\delta g_{Lb}$, 
but this can be compensated by an enhancement $(g_R/g_L)^2\sim
\cot^2\theta_R$ in the right-handed gauge couplings.

It would be impossible to present a complete one-loop calculation 
for $Z \rightarrow b\bar{b}$ in the general case of Section~\ref{secmods} because 
we do not have sufficient information at present to determine all the 
parameters in those models. At the same time, we are interested 
in exploring the idea of a potentially strong right-handed interaction 
affecting the $b$ and $t$-quarks more than we are interested in the 
specific details of the models in Section~\ref{secmods}. 
For this reason, we consider a slightly 
simpler calculation that has the ingredients we need. 
First, we will only concern ourselves with the 
one-loop corrections that are enhanced by $(g_R/g_L)^2$ with respect to 
one-loop electroweak corrections. Second, we will require that there 
be no $Z-Z^\prime$ nor $W-W^\prime$ mixing in the model. 
Finally, we will treat all standard model fermions 
as massless except for the top-quark.

\subsection{Model with no tree-level mixing}

To eliminate the tree-level mixing in the models of Section~\ref{secmods} 
in a simple manner we first require $v_2 = 0$ in Eq.~\ref{massmat}. 
This immediately makes $\xi_W =0$ and allows us to simplify the notation 
by calling the remaining VEV in $\phi$ $v \equiv v_1$. We further 
make $\xi_Z =0$ at tree-level by imposing the condition,
\begin{equation}
v_L = v \cot\theta_R 
\end{equation}
in Eq.~\ref{massmat}. The parameter $\xi_Z$ describing the 
$Z-Z^\prime$ mixing is the only one (beyond those already appearing in 
the standard model) that enters the result for the 
$Z \rightarrow b \bar{b}$ partial width at tree-level. As such, it is the 
only new parameter that needs to be defined at one-loop in our 
calculation of $Z \rightarrow b \bar{b}$, and we will return to this 
point at the end of the section.

In the simplified model, the gauge boson masses become,
\begin{eqnarray}
M_W^2 = {g_L^2 \over 2} \left( v_L^2 + v^2 \right)
 =  {g_R^2 \over 2 \tan^2\theta_W}v^2 &,& 
M_Z^2 =  {M_W^2 \over \cos^2\theta_W} \nonumber \\
M_{W_R}^2 = {g_R^2 \over 2} \left( v_R^2 + v^2 \right) &,& 
M_{Z_R}^2 = {g_R^2 \over 2} \left( {v_R^2\over \cos^2\theta_R} + v^2 \right) 
\label{simpmass}
\end{eqnarray}
so that $M_{W_R}\approx M_{Z_R}$ for large $\cot\theta_R$ (and equal to 
$M_{W^\prime}$ and $M_{Z_\prime}$ since in this case there is no mixing). 
Recalling the relation, $g_R \sin\theta_R = g_L \tan\theta_W$, 
we see that this sector of the model is characterized by the two 
ratios of vevs
\begin{eqnarray}
{v_L \over v} = \cot\theta_R \, \, , \,\,
{v_R \over v} \approx {M_{W_R}\over M_W \tan\theta_W} 
\label{ratiosvevs}
\end{eqnarray}
where the last expression follows for $v_R/v >> 1$. It will also 
be convenient to define $x_v = v^2 /( v^2 + v_R^2)$. 

In the approximation $m_b=0$, the Yukawa Lagrangian necessary to 
generate the top-quark mass is given by,
\begin{eqnarray}
{\cal L}_Y = - \kappa 
\left ( \begin{array}{lr} \bar{t}_L ~~\bar{b}_L \end{array} \right)
\phi
\left( \begin{array}{l}
t_R \\
b_R
\end{array}
\right ) \, + \, {\rm h.c.}
\end{eqnarray}
All the couplings in this Yukawa potential can thus be written 
in terms of $m_t$ and $v$ as in Eq.~\ref{yukawa}. 

\subsection{Loops with right-handed gauge bosons} 

\begin{figure}[!htb]
\begin{center}
\includegraphics[width=12cm]{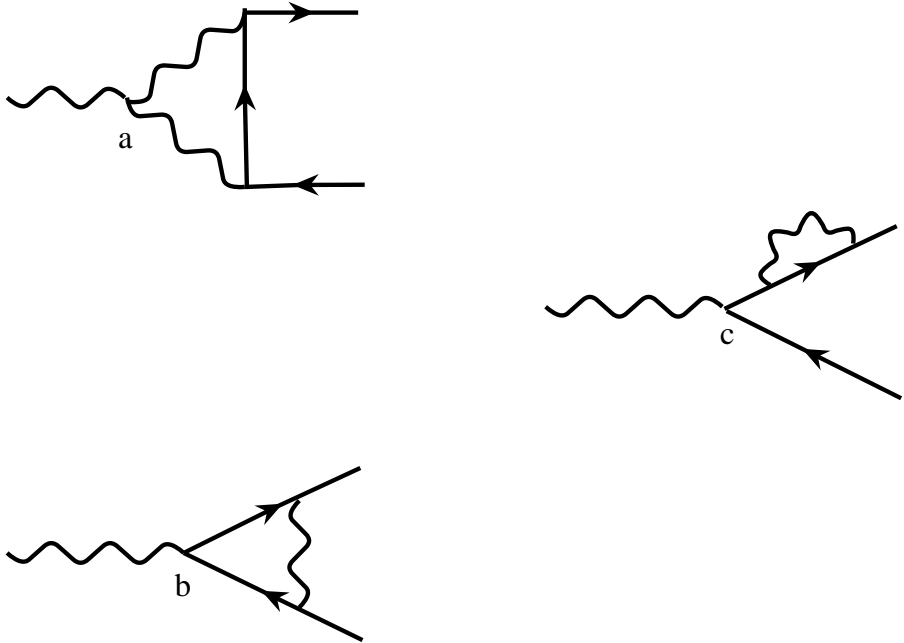}
\end{center}
\caption{Unitary gauge Diagrams for $Z\rightarrow b\bar{b}$ that do not 
involve scalars.}
\label{diagsw}
\end{figure}

We are ready to calculate the one-loop corrections to $\delta g_{Rb}$ that are 
enhanced by $\cot^2\theta_R$. We start by considering the diagrams in 
Figure~\ref{diagsw} that do not involve scalar mesons in the loop. 
We work in unitary gauge with the vertices 
given in the Appendix, and we use dimensional regularization with  
the notation,
\begin{equation}
{1\over \epsilon} = {2\over 4-n}-\gamma + \log(4\pi)-\log\;\mu^2 \, .
\end{equation}

Relegating details for each diagram to the Appendix, we can 
write a simple analytical result for the sum 
of the diagrams in Figure~\ref{diagsw} in the limit 
$M_Z=0$,
\begin{equation}
\left(\delta g_{Rb}\right)_{GB} = {g_R^2 \over 16 \pi^2}{M_t^2\over M_{W_R}^2}
{1\over 4}\left({1\over \epsilon}-{7\over 2}-
\log\left({ M_t^2\over \mu^2}\right)-3
\log\left({ M_t^2\over M^2_{W_R}}\right)\right)
\label{wloopsr}
\end{equation}
Later on we will show numerical results for $M_Z\neq 0$. Unlike the 
counterpart of this calculation in the standard model, Eq.~\ref{smf}, 
Eq.~\ref{wloopsr} is divergent. This indicates the presence of 
additional contributions to this  process in our model. 

There are two additional diagrams of the form of Figure~\ref{diagsw}
that give corrections to the right-handed 
coupling and that are enhanced by $g_R^2$. They look like the diagrams 
(b) and (c)  with an exchange of a $Z_R$ (and therefore $b$-quarks in the 
intermediate lines). Both of these turn out 
to be finite and their finite parts precisely cancel each other out in the 
$M_Z=0$ limit.

\subsection{Loops with Scalars}

\begin{figure}[!htb]
\begin{center}
\includegraphics*[width=12cm]{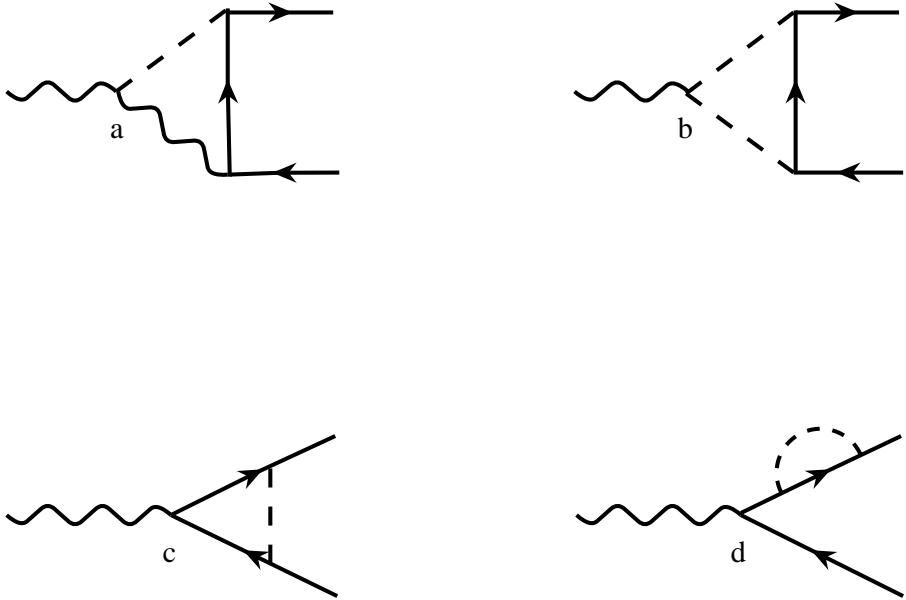}
\end{center}
\caption{$Z \rightarrow b\bar{b}$ one-loop diagrams involving scalars}
\label{hdiags}
\end{figure}

We consider next the contributions from diagrams in which scalars 
appear in the loops, as in Figure~\ref{hdiags}.

The finite part that results from these diagrams is model dependent. In 
particular it depends on the details of the scalar potential, which we 
have not specified, and which determines the masses of the physical 
scalar and pseudo-scalar mesons present in the model. We are only interested 
here in estimating the size of the vertex corrections in Figure~\ref{diagsw}, 
and wish to consider the diagrams in Figure~\ref{hdiags} only insofar as they 
are needed to render the  result finite. For this purpose it is sufficient to 
identify a basis for the scalars that is orthogonal to the would-be-Goldstone 
bosons that give the gauge bosons their mass. We consider all physical 
scalars to be degenerate and to have a large mass, of order 
$M_{W_R}$. With details relegated to the Appendix, we find that the sum of these 
diagrams contributes the following terms that are enhanced by 
$\cot^2\theta_R$,
\begin{eqnarray}
\left(\delta g_{Rb}\right)_{S} &=& 
{1\over 16 \pi^2}\left({m_t\over v}\right)^2
\left(-{1\over 2}x_v(1-x_v)\right)\left({1\over \epsilon}-\log\left({M_H^2\over
      \mu^2}\right) +{5\over 2} \right) \nonumber \\
&=& {g_R^2\over 16 \pi^2}{M_t^2\over M^2_{W_R}}{1\over 4}
\left( -{1\over \epsilon}+\log\left(
{M_H^2\over \mu^2}\right)-{5\over 2}\right)
\label{hdiagsr}
\end{eqnarray} 
Notice that the left-handed coupling $\delta g_{Lb}$ does not receive 
corrections from the sum of diagrams in Figure~\ref{hdiags}. 
With this result, Eq.~\ref{hdiagsr}, we find that 
the divergent terms precisely cancel the 
left-over ones from the gauge boson sector in Eq.~\ref{wloopsr} 
leaving us with a finite 
answer.

\subsection{Renormalization and $Z-Z^\prime$ Mixing}

\begin{figure}[!htb]
\begin{center}
\includegraphics*[width=12cm]{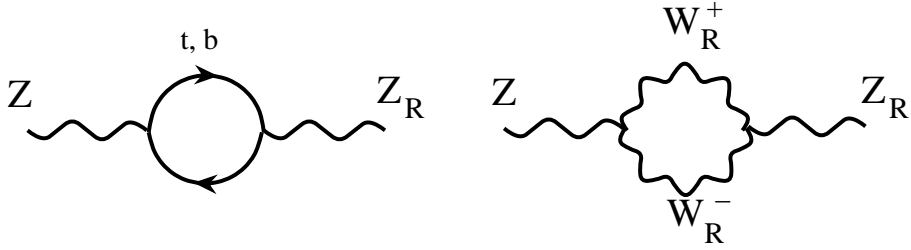}
\end{center}
\caption{$Z-Z^\prime$ Mixing at one-loop}
\label{zzrmix}
\end{figure}

Finally we comment on the renormalization scheme used. At tree level, 
the $Z\rightarrow b \bar{b}$ decay width (or $A^b_{FB}$) takes the 
same value as in the standard model in the absence of $Z-Z^\prime$ mixing. 
We can therefore express it in terms of the input parameters $G_F$, the 
physical $Z$ mass and $\alpha(M_Z)$ as is usually done for the standard 
model case. It is clear from the vertices given in the Appendix, 
Section~\ref{secfr}, that none of these quantities receives one-loop 
corrections that are enhanced by $\cot^2\theta_R$. The only input 
parameter that receives enhanced corrections is the $Z-Z^\prime$ 
mixing angle $\xi_Z$ through diagrams such as those in Figure~\ref{zzrmix}. 
These diagrams (and a few others), have an enhancement of $\cot\theta_R$ 
through the $Z^\prime t\bar{t}$ or $Z^\prime b\bar{b}$ coupling in the 
first case and through the $W^+_R W^-_R Z^\prime$ coupling in the second 
case. When the $Z^\prime$ line is connected to $b\bar{b}$, a second 
$\cot\theta_R$ factor is picked up leading to corrections 
in $Z \rightarrow b \bar{b}$ that are enhanced by $\cot^2\theta_R$. 
In view of this, our simplest option is to adopt a renormalization scheme 
in which $\xi_Z \equiv 0$ at one-loop. That is, we absorb the corrections from 
Figure~3 into the definition of $\xi_Z$. This completes the discussion of 
all the one-loop corrections needed to yield a finite $\delta g_{Rb}$ and 
we now turn our attention to its possible size.

\subsection{Numerical Results}

Adding the results from all diagrams discussed above, we find 
in the $M_Z=0$ limit, 
\begin{equation}
\delta g_{Rb}= 
{g_R^2\over 16 \pi^2}{M_t^2\over M^2_{W_R}}
{1\over4}\left(-6+3\log\left({M^2_{W_R}\over M_t^2}\right)
+\log\left({M_H^2\over M_t^2}\right)\right).
\end{equation}
To illustrate the magnitude of this correction consider the case,
\begin{equation}
\cot\theta_R = {v_L\over v} \approx {v_R \over v} \sim 10
\label{casea}
\end{equation}
which implies 
\begin{equation}
{g_R^2\over M^2_{W_R}} \approx {g_L^2\over M^2_W}
\end{equation}
and therefore
\begin{equation}
\delta g_{Rb} = {\sqrt{2} G_F M_t^2 \over 8 \pi^2}
\left(-3+{3\over 2}\log\left({M^2_{W_R}\over M_t^2}\right)
+{1\over 2}\log\left({M_H^2\over M_t^2}\right)\right). 
\end{equation}
This is to be compared with the corresponding correction to 
$\delta g_{Lb}$ in the standard model which is given by \cite{Altarelli:hv}
\begin{equation}
\delta g_{Lb} =  {\sqrt{2} G_F M_t^2 \over 8 \pi^2}.
\label{smf}
\end{equation}
This shows that with large $\cot\theta_R$ as in Eq.~\ref{casea}, 
$\delta g_{Rb}$ in our model is of the same order as the 
one-loop correction to $\delta g_{Lb}$ proportional to $M_t^2$ in the 
standard model.

\begin{figure}[!htb]
\begin{center}
\includegraphics*[width=10cm]{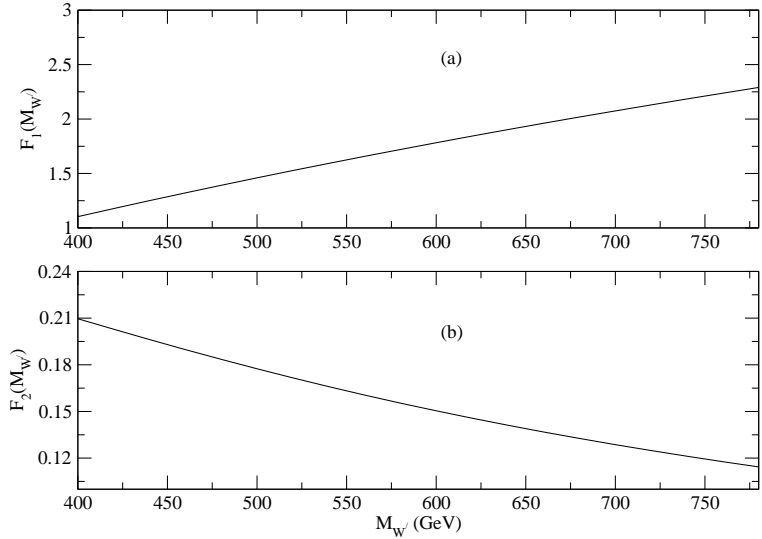}
\end{center}
\caption{Form factors $F_1(M_{W_R})$ and $F_2(M_{W_R})$ evaluated
numerically to include kinematic effects from a non-zero $M_Z$ 
for fixed $M_H=700$~GeV.}
\label{numres}
\end{figure}

In order to include kinematic effects from $M_Z \neq 0$, 
we compute the integrals over Feynman parameters numerically. 
It is convenient to present the result in the form
\begin{equation}
\delta g_{Rb} = {\sqrt{2} G_F M_t^2 \over 8 \pi^2} F_1(M_{W_R})
\end{equation}
for the case $g_L^2/M_W^2 \approx g_R^2/M^2_{W_R}$. We show 
$F_1$ in Figure~\ref{numres}~a. In the more general case it is 
convenient to write
\begin{equation}
\delta g_{Rb} = {g_R^2 \over 32 \pi^2} F_2(M_{W_R})
\end{equation}
and we show $F_2$ in Figure~\ref{numres}~b. 
\vspace{0.5in}
These results indicate that a contribution to $\delta g_{Rb}$ at the $\%$ level 
is possible in models with $\cot\theta_R \sim 10$.

\section{Constraints from LEP-II} \label{seclep2}

The mass of additional $Z^\prime$ gauge bosons that occur in many models 
is constrained to be larger than about 500~GeV \cite{Hagiwara:fs}. 
These bounds arise mostly from 
processes involving four first or second generation fermions and do not apply
to non-universal $Z^\prime$ gauge bosons that couple strongly to third
generation fermions but very weakly to first and second generation fermions. 
Roughly speaking, when a $Z^\prime$ like this is exchanged in a  
process such as $b\bar{b} \rightarrow b\bar{b}$ it generates an 
amplitude of order electroweak strength times $\cot^2\theta_R$. For 
the models that we have in mind $\cot\theta_R \sim 10$ this can be a 
very significant enhancement. On the other hand when the same $Z^\prime$ 
is exchanged between fermions of the first two generations, in processes 
such as $u\bar{u} \rightarrow u \bar{u}$, it generates an amplitude of 
electroweak strength times $\tan^2\theta_R$ which is drastically 
suppressed. 

The best bounds one can have at present on such a $Z^\prime$ come 
from a process in which a first or second generation fermion pair 
produces a $b\bar{b}$  pair. Since $b\bar{b}$ production in hadron 
colliders is mostly a strong interaction process, the most promising 
reaction to constrain our $Z^\prime$ is $e^+e^- \rightarrow b\bar{b}$ 
studied at LEP-II. Notice that for a process such as this one, the 
exchange of a $Z^\prime$ results in a correction of electroweak 
strength, suppressed only by the mass of the $Z^\prime$. 
The cross-section for this  process is largely independent of 
the value of $\cot\theta_R$. At leading order, $\cot\theta_R$ only appears 
through the width of the $Z^\prime$ that one must include in the 
propagator for $s$-channel exchange.

In this Section we use the LEP-II data on $e^+ e^- \rightarrow b\bar{b}$ and 
$e^+ e^- \rightarrow \tau^+ \tau^-$ to constrain the mass of these non-universal 
$Z^\prime$ gauge bosons. 
The calculation is performed numerically using the program COMPHEP
\cite{Pukhov:1999gg} with the following strategy. We use COMPHEP to calculate 
tree level cross-sections for  $e^+ e^- \rightarrow f\bar{f}$ at LEP-II 
energy both for the standard model and for the standard model plus the 
$Z^\prime$ of Section~II. We then use these cross-sections to construct the ratios
$R_b/(R_b)_{SM-tree}$ and similarly for $A^b_{FB}$. We then compare these ratios 
to the corresponding ratios $(R_b)_{EXP}/(R_b)_{SM}$ where $(R_b)_{EXP}$ 
are the averages of LEP-II measurements as reported in
Ref.~\cite{Abbaneo:2001ix} and $(R_b)_{SM}$ is the full 
standard model expectation computed with ZFITTER as reported in Table~8.7 of 
Ref.~\cite{Abbaneo:2001ix}. Schematically for the cross-section,
\begin{eqnarray}
{\sigma_{Theory} \over \sigma_{SM} } &=& 
{\sigma_{SM-tree} + \sigma_{SM-loop} + \sigma_{Z^\prime} \over 
\sigma_{SM-tree} + \sigma_{SM-loop}} \nonumber \\
&\approx & 
{\sigma_{SM-tree} + \sigma_{Z^\prime} \over \sigma_{SM-tree}} 
\, = \, 1 + {\sigma_{Z^\prime} \over \sigma_{SM-tree}}.
\end{eqnarray}
In this way the error that results from our using only the tree-level 
result for the standard model prediction from COMPHEP becomes higher 
order in our comparison with data.

\begin{figure}[!htb]
\begin{center}
\includegraphics*[width=12cm]{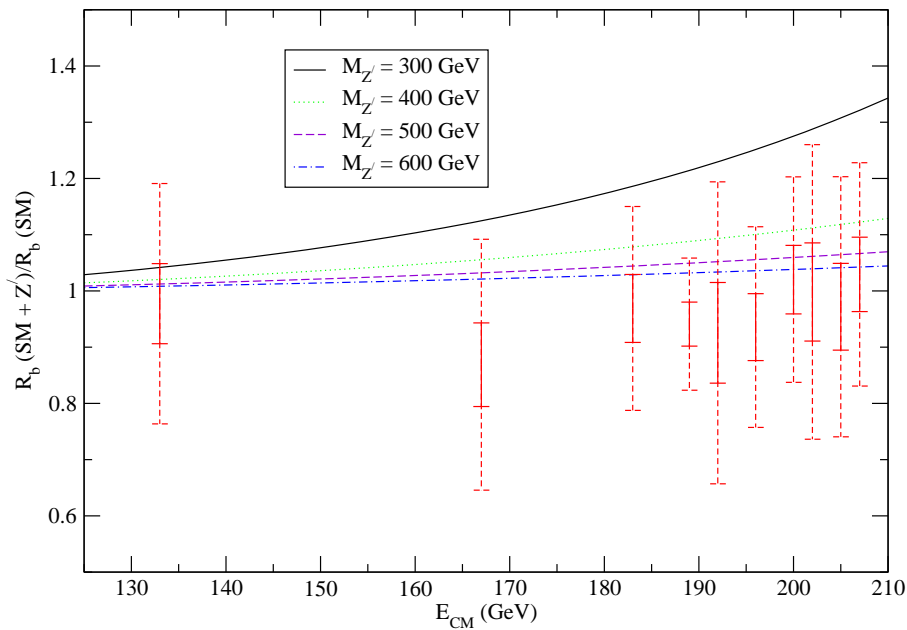}
\end{center}
\caption{$R_b$ at LEP-II energies for $\cot\theta_R = 15$ with 
no $Z-Z^\prime$ mixing. The different curves correspond to $M_{Z^\prime}$ 
of 300, 400, 500 and 600~GeV. The data points from Ref.~\cite{Abbaneo:2001ix} 
are shown with their $1$-$\sigma$ and $3$-$\sigma$ error bars.}
\label{rbfig}
\end{figure}

In models with $Z-Z^\prime$ mixing, we need to remove the enhanced 
coupling $Z^\prime \tau^+\tau^-$ as discussed in the previous section. 
In that case the only relevant LEP-II process to bound the $Z^\prime$ 
is $e^+ e^- \rightarrow b\bar{b}$. There are two observables that can be 
used: $R_b$ and $A^b_{FB}$. In Figure~\ref{rbfig} we show 
$R_b/(R_b)_{SM}$ for different values of $M_Z^\prime$. In this figure 
we have assumed no mixing ($\xi_Z=0$) and used $\cot\theta_R = 15$ 
\footnote{recall that the dependence of this amplitude on $\cot\theta_R$ is only 
through the width of the $Z^\prime$}.  The LEP-II data points are 
shown with their $1$-$\sigma$ and $3$-$\sigma$ error bars. It is evident 
already from this figure that $M_{Z^\prime}$ will be constrained to 
be larger than about 500~GeV.

\begin{figure}[!htb]
\begin{center}
\includegraphics*[width=12cm]{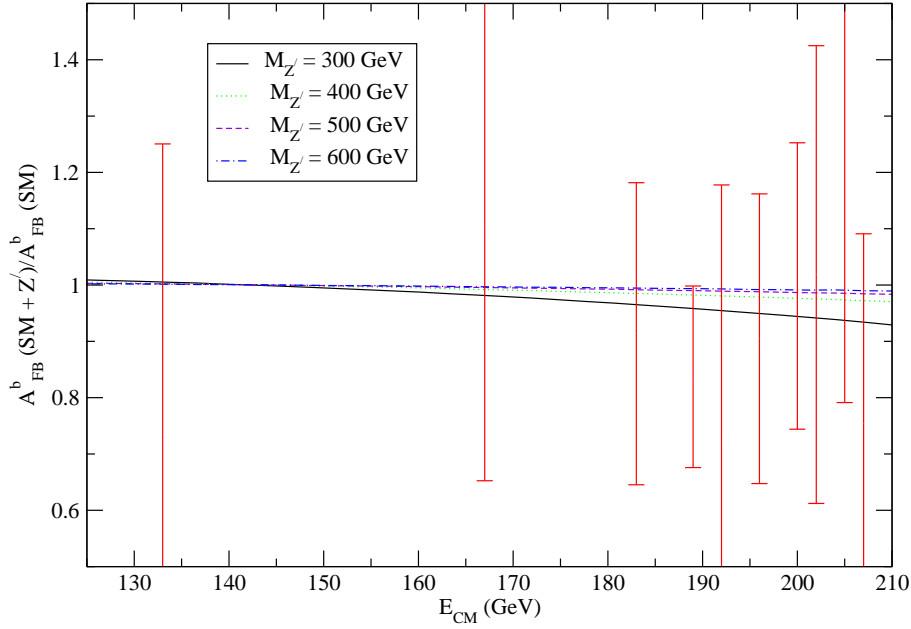}
\end{center}
\caption{Same as Figure~\ref{rbfig} for $A^b_{FB}$. Only the $1$-$\sigma$ 
error bars are shown for the data points from Ref.~\cite{Abbaneo:2001ix}.}
\label{abfig}
\end{figure}

In Figure~\ref{abfig} we show similar results for the forward-backward 
asymmetry. It is evident from this figure that $A^b_{FB}$ does not 
constrain the $Z^\prime$ as much as $R_b$ does due to its larger 
experimental error (in this case we only show the $1$-$\sigma$ error-bars).

\begin{figure}[!htb]
\begin{center}
\includegraphics*[width=12cm]{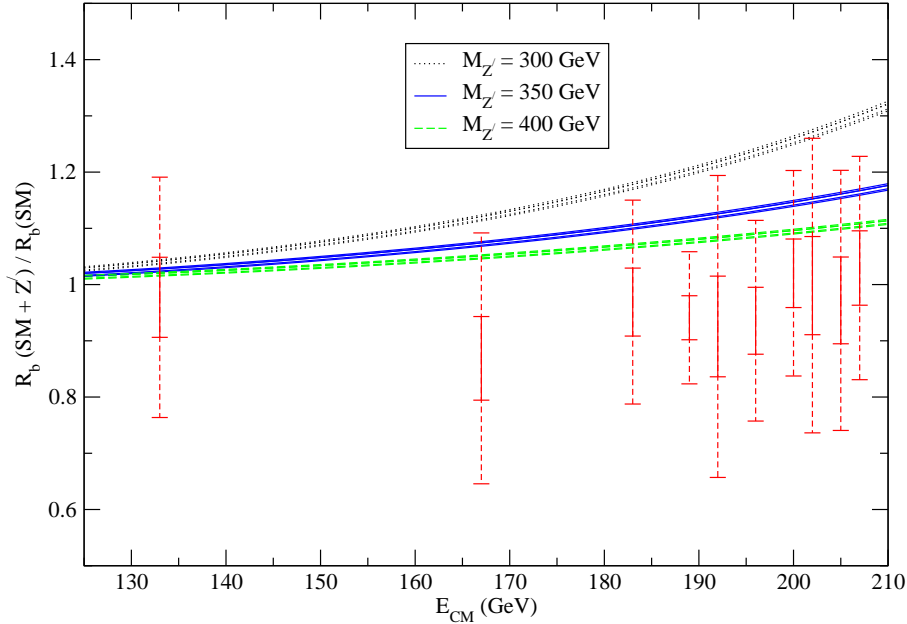}
\end{center}
\caption{Same as Figure~\ref{rbfig} but allowing for $Z-Z^\prime$ mixing. The 
bands shown correspond to $\cot\theta_R \xi_z$ ranging from 0 to $\pm
0.08$ with $\cot\theta_R = 15$. For each mass, the upper end of the band 
corresponds to 
$\xi_Z = -0.005$ and the lower end of the band corresponds to $\xi_Z =
0.005$. Once again the data points are from Ref.~\cite{Abbaneo:2001ix}.}
\label{rbfigm}
\end{figure}

As discussed in Section~\ref{secmods}, it is possible to allow $Z-Z^\prime$ mixing 
in models where the couplings to $\tau^\pm$ are not enhanced. We illustrate 
the effect of including this mixing in Figure~\ref{rbfigm}. For each 
value of $M_{Z^\prime}$, we have allowed $\xi_Z$ to vary between zero and 
$\xi_Z = \pm 0.08 /\cot\theta_R$, the value required to fit $A^b_{FB}$ 
from LEP-I. We see that mixing is a small effect on $R_b$ at LEP-II 
energies. 

\begin{figure}[!htb]
\begin{center}
\includegraphics*[width=12cm]{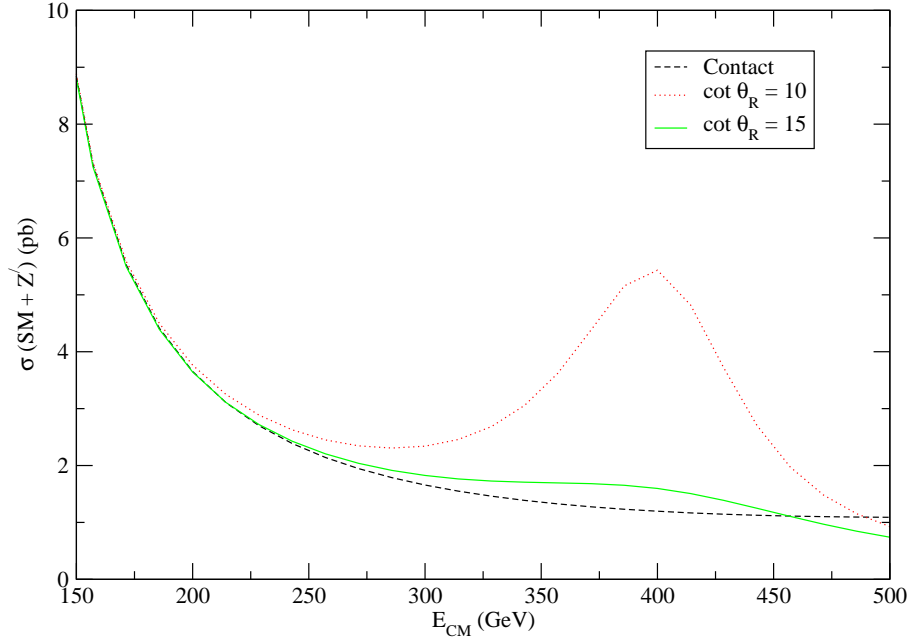}
\end{center}
\caption{Cross-section for $e^+e^-\rightarrow b\bar{b}$ with 
$M_{Z^\prime}=400$~GeV for $\cot\theta_R = 10,\, 15$ and for a 
contact interaction approximating the $Z^\prime$ exchange amplitude.}
\label{mzp400}
\end{figure}

In Figure~\ref{mzp400} we illustrate the effect of varying 
$\cot\theta_R$ and thus the $Z^\prime$ width for $M_{Z^\prime}=400$~GeV. 
We use values $\cot\theta_R = 10,\, 15$ and also show the result of 
approximating the $Z^\prime$ exchange with a contact interaction. The 
results illustrate that below the resonance, the bound on the $Z^\prime$ 
mass becomes slightly tighter for narrower resonances (smaller 
$\cot\theta_R$). In our model, the interference between $Z$ and 
$Z^\prime$ exchange amplitudes is always constructive in the energy 
region between the two resonances. The figure also illustrates that for 
a $Z^\prime$ as light as 400~GeV, a contact interaction is a reasonable 
approximation for effects at LEP-II energies. We shall use this later 
when comparing our bounds with those extracted by the LEP-II analysis 
group for contact interactions.

\begin{figure}[!htb]
\begin{center}
\includegraphics*[width=12cm]{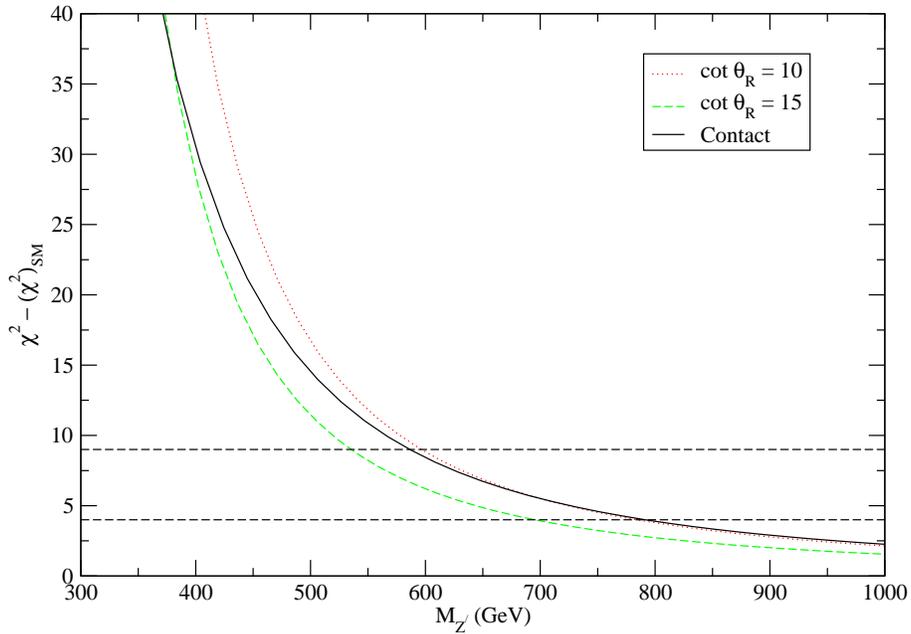}
\end{center}
\caption{$\chi^2-\chi^2_{SM}$ for fits to $R_b$ at LEP-II energies 
as a function of $M_{Z^\prime}$ setting $\xi_Z=0$.}
\label{chicomp}
\end{figure}

To quantify the bounds on the $Z^\prime$ mass we construct a 
$\chi^2$ for a fit to LEP-II data with our model. In Figure~\ref{chicomp} 
we present this result after subtracting the $\chi^2$ from a 
standard model fit (using the ZFITTER results quoted in
Ref.~\cite{Abbaneo:2001ix}). Once again we show the three cases 
$\cot\theta_R = 10,15$ and a contact interaction approximation. 
It is important to notice, for example in Figure~\ref{rbfig}, that 
the LEP-II data are consistently below the standard model prediction. 
This, combined with the fact that the interference between the $Z$ and 
$Z^\prime$ amplitudes in our model is always constructive in this energy region, 
implies that the standard model is always a better fit than any of our 
$Z^\prime$ models. If we require that the new model not deviate from the 
standard model by 2(3) standard deviations, we can place the bounds 
$M_{Z^\prime} > 700 (540)$~GeV for $\cot\theta_R=15$.
Given that the LEP-II data is consistently below the standard
model expectation, 
it is conceivable that there is some common systematic 
error not accounted for in the quoted error bars. 
To account for this possibility, we naively rescale 
the data by a common factor in such a way as 
to minimize the $\chi^2$ of the standard model fit. Doing this 
results in lower bounds on the $Z^\prime$ mass. For example, 
for $\cot\theta_R =10$ the 2(3) sigma bounds move from 780~(595)~GeV 
to 530~(460)~GeV.

We can also use the contact interaction approximation to bound the 
$Z^\prime$ mass. The correspondence to Table~8.12 of Ref.\cite{Abbaneo:2001ix} 
is (for no mixing),
\begin{equation}
M_{Z^\prime} = \sqrt{\eta}{g_L\tan\theta_W \over 4\sqrt{\pi}} \Lambda
\end{equation}
Our model of Section~II generates both a $LR$ contact interaction 
with $\eta_{LR} = 1$ and a $RR$ contact interaction with $\eta_{RR} = 2$. 
For constructive interference we thus infer the 95\% confidence level 
bounds $M_{Z^\prime}> 544$~GeV from the $RR$ interaction and 
$M_{Z^\prime}> 275$~GeV from the $LR$ interaction.

From all this we conclude that in all cases the 
$Z^\prime$ is already restricted to be heavier than about $500$~GeV 
by LEP-II data.

\begin{figure}[!htb]
\begin{center}
\includegraphics*[width=12cm]{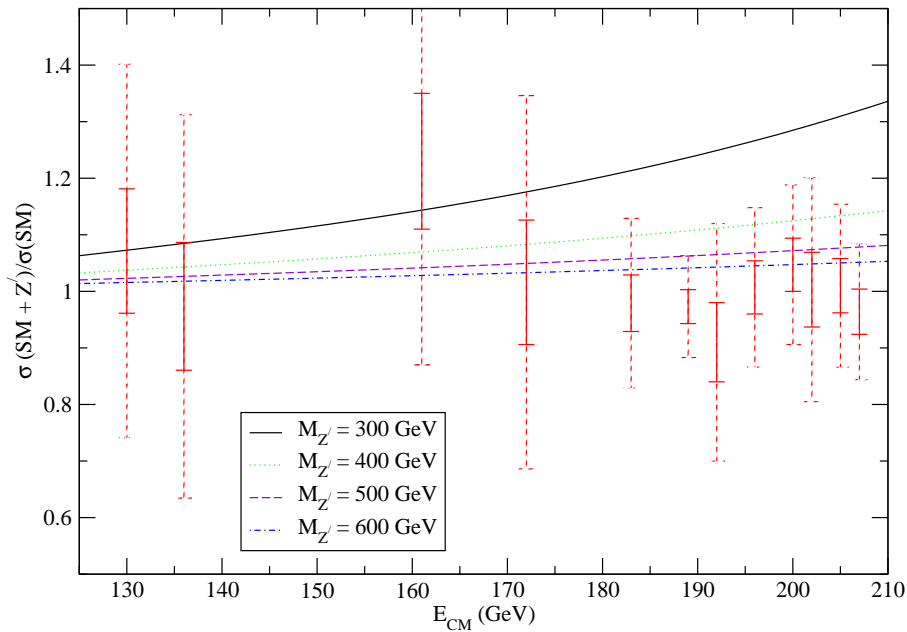}
\end{center}
\caption{$\sigma(e^+e^-\rightarrow \tau^+\tau^-)$ for $\cot\theta_R = 15$. The
data points are from Ref.~\cite{Abbaneo:2001ix}.}
\label{sigttfig}
\end{figure}

Finally, since models without mixing also allow large couplings to the 
$\tau$-lepton, we show in Figure~\ref{sigttfig} the 
cross-section for $e^+e^-\rightarrow \tau^+\tau^-$ at LEP-II. 
A calculation of the $\chi^2$ for the fit in this case indicates that 
the bounds on the $Z^\prime$ mass are slightly higher than those obtained 
from studying $R_b$, but not significantly so.

\clearpage

\section{Conclusions}

Motivated by the 3-$\sigma$ discrepancy between the standard model 
prediction and the measured forward-backward asymmetry $A^b_{FB}$ at 
the $Z$ peak we have studied models which can generate a sufficiently 
large $\delta g_{Rb}$ through new non-universal right-handed gauge 
interactions. 

One possible mechanism to generate this $\delta g_{Rb}$ is the mixing 
of the $Z$ with a $Z^\prime$. We had already discussed a model like 
this in Ref.~\cite{He:2002kn}. In this paper we have illustrated 
several variations on that model that are also renormalizable and anomaly 
free. At the cost of introducing additional fermions, we showed two 
models that produce the required $\delta g_{Rb}$ while satisfying the 
LEP constraints on $\delta g_{R\tau}$. We have also indicated how it 
is possible to modify these models so that they are not constrained 
by $b\rightarrow s \gamma$. 

We have identified a second mechanism to generate $\delta g_{Rb}$ 
even in cases with no $Z-Z^\prime$ mixing. This occurs in models with 
an $SU(2)_R$ triplet of gauge bosons at one-loop, and can give rise 
to $\delta g_{Rb}$ at the $1\%$ level. By itself, this mechanism is 
not sufficient to explain the full $\delta g_{Rb} \sim 0.04$ favored by the 
data. The simple model used to illustrate this effect, provides an example 
of a renormalizable model that can give rise to relatively large new 
interactions involving only the $b$ and $t$ quarks while respecting 
low energy constraints. 

Finally we have used the LEP-II data for the process $e^+e^- \rightarrow b
\bar{b}$ to place bounds on the mass of the $Z^\prime$ in our models. 
This is a $Z^\prime$ that couples weakly (by a factor $\tan\theta_R\sim 1/10$ 
weaker than standard electroweak couplings) to fermions of the first two 
generations. For this reason, standard bounds on $Z^\prime$ gauge bosons 
do not apply. We find that the LEP-II data constrains it to be heavier 
than about $500$~GeV in all cases. 

The contribution of the new gauge bosons in our models to the process 
$e^+e^- \rightarrow b \bar{b}$, is of electroweak strength because the 
enhancement in the $Z^\prime b \bar{b}$ coupling is compensated by the 
suppression in the $Z^\prime e^+e^-$ coupling. In this way, our model is 
an example of a kind of new interactions that will only show their full 
strength in processes involving four third generation fermions. It may be 
possible for the LHC to study certain processes of this type, and we are 
currently investigating this possibility. 

\noindent {\bf Acknowledgments}$\,$    
The work of X.G.H. 
was supported in part by National Science Council under grants NSC
91-2112-M-002-042, and in part by the Ministry of
Education Academic Excellence Project 89-N-FA01-1-4-3.
The work of G.V. was supported in part by DOE under 
contract number DE-FG02-01ER41155.   

\appendix

\section{Vertices and One-Loop Results}

\subsection{Basic conventions}

The general conventions adopted are:
\begin{eqnarray}
g_{Lt} &=& 1 - {4\over 3}\sin^2\theta_W,  \,\,
g_{Rt} =  - {4\over 3}\sin^2\theta_W, \nonumber \\
g_{Lb} &=& -1 + {2\over  3}\sin^2\theta_W, \,\,
g_{Rb} =  {2\over 3}\sin^2\theta_W, \nonumber \\
\Gamma_\mu &=& -i {g \over 2 \cos\theta_W} \gamma_\mu \left(
(g_{Lb} + \delta g_{Lb})P_L + (g_{Rb} + \delta g_{Rb})P_R \right) \nonumber \\
{1\over \epsilon} &=& {2\over 4-n}-\gamma + \log(4\pi)-\log\;\mu^2 
\label{conventions}
\end{eqnarray}

\subsection{Scalar Sector}

We start with the following parameterization for the scalars,
\begin{eqnarray}
&& H_L = \left ( \begin{array}{c}
{h_L-i\phi_L \over \sqrt{2} }+ v_L\\
\phi^-_L
\end{array}
\right ),\;\;
H_R = \left ( \begin{array}{c}
{h_R-i\phi_R \over \sqrt{2} }+ v_R\\
\phi^-_R
\end{array}
\right ), \nonumber \\
&& \phi = \left ( \begin{array}{cc}
{h_1-i\phi^0_1 \over \sqrt{2} }+ v_1 & \phi^+_1   \\
\phi^-_2  & {h_2-i\phi^0_2 \over \sqrt{2} }+ v_2 
\end{array}
\right )
\label{scalpara}
\end{eqnarray}
This parameterization contains both the would-be Goldstone-bosons that 
give mass to the $W,Z,W_R,Z_R$ and the remaining physical scalar (or 
pseudo-scalar) particles.

Since we do not specify a scalar potential, we cannot identify the 
scalar mass eigenstates. Rather we work with a basis of physical 
scalars chosen to be orthogonal to the would-be Goldstone-bosons, 
under the simplifying assumption that they all have the same large mass. 
The physical scalars defined this way are,
\begin{eqnarray}
H_1^\pm &=& {1\over \sqrt{v^2+v_L^2}}(-v \phi_L^\pm +v_L \phi_2^\pm) 
\nonumber \\
H_2^\pm &=& {1\over \sqrt{v^2+v_R^2}}(-v \phi_R^\pm +v_R \phi_1^\pm) 
\nonumber \\
H_2^0 &=& {x\over \sqrt{1+x^2}}(-\sin\theta_R \phi^0_L +{1\over x}\phi^0_R
+\cos\theta_R \phi^0_1)
\nonumber \\
H_1^0 &=& \phi^0_2 
\label{scalarbasis}
\end{eqnarray}
with $x = v_R/ (v \cos\theta_R)$ and with 
$H_{1,2}^0$ being neutral pseudo-scalars. The only neutral scalar 
that enters the calculation is  the original $h_2$. 
Working with this basis we can identify the 
divergent contributions arising from diagrams with scalar exchange (they 
are independent of scalar masses). The finite contributions that depend on 
the masses of the different scalars can only be obtained 
after fixing all their masses.

\subsection{Feynman Rules}\label{secfr}

In unitary gauge, the vertices 
$\gamma W^+_L W^-_L$, $\gamma \gamma W^+_L W^-_L$, 
$\gamma Z W^+_L W^-_L$, $Z W^+_L W^-_L$ and $ZZ W^+_L W^-_L$ 
are as in the standard model. The analogous vertices with 
$W^\pm_R$ taking the place of $W^\pm_L$ can be obtained by 
multiplying the corresponding vertex with $W^\pm_L$  by a factor 
of $(-\tan^2\theta_W)$ for each $Z$. In particular, these vertices 
are {\it not} enhanced by $g_R/g_L$. There are no ``mixed'' vertices 
with one $W_L$ and one $W_R$. For vertices
involving a $Z_R$, the $Z_R W^+_R W^-_R$ vertex is given by
$\tan\theta_W\cot\theta_R/\cos\theta_W$ times the corresponding vertex 
for $Z W^+_L W^-_L$ and is, therefore, enhanced with respect to the latter 
by a factor of $\cot\theta_R$. Finally the $Z Z_R W^+_R W^-_R$ vertex 
can be obtained by multiplying the $ZZ W^+_L W^-_L$ vertex by 
$-\tan^3\theta_W\cot\theta_R/\cos\theta_W$. 

The Feynman rules for couplings of gauge bosons to fermions 
are already given in Eqs.~\ref{cccoup}~and~\ref{neucoupl}. 
The Feynman rules involving 
the scalars which couple to top and bottom proportionally to the top-quark 
mass can be extracted from the Yukawa Lagrangian:
\begin{eqnarray}
{\cal L}_Y &=&  -m_t \left[ \bar{t}t + {1\over \sqrt{2}v} 
(\bar{t}t h_1 + \bar{b}b h_2) -{i\over \sqrt{2}v}
(\bar{t}\gamma_5 t \phi^0_1 + \bar{b}\gamma_5 b \phi^0_2) \right.
\nonumber \\
&+& \left. {1 \over v} \left( \bar{b}_L t_R \phi^-_2 +  \bar{b}_R t_L \phi^-_1
+ \bar{t}_L b_R \phi^+_1 +  \bar{t}_R b_L \phi^+_2 \right) \right]
\label{yukawa}
\end{eqnarray}
In terms of the physical scalars defined as in Eq.~\ref{scalarbasis} the  
couplings we need become,
\begin{eqnarray}
{\cal L}_Y &=&  -{m_t \over v} \left[
\cos\theta_R \left(\bar{b}P_R t H_1^- + \bar{t} P_L b H^+_1\right) 
+ {v_R \over \sqrt{v_R^2+v^2}}
\left(\bar{b}P_L t H_2^- + \bar{t} P_R b H^+_2\right) \right.
\nonumber \\
 &+& \left. {1\over \sqrt{2}}\bar{b} \left(h_2 - i H^0_1\gamma_5\right) b \right].
\label{yukphy}
\end{eqnarray}
The vertices of the form $Z HH$ are obtained from the Lagrangian
\begin{eqnarray}
{\cal L} &=& {i g_L \over 2 \cos\theta_W}
\left(2\sin^2\theta_W-{v_R^2\over v^2+v_R^2}\right)
Z^\mu\left(H_2^-\partial_\mu H_2^+ - H_2^+\partial_\mu H_2^-\right) \nonumber \\
&&{-i g_L \over 2 \cos\theta_W}\cos 2\theta_W
Z^\mu\left(H_1^-\partial_\mu H_1^+ - H_1^+\partial_\mu H_1^-\right) \nonumber \\
&+&{ g_L \over 2 \cos\theta_W}
Z^\mu\left(h_2\partial_\mu H_1^0 - H_1^0\partial_\mu h_2\right) .
\end{eqnarray}
Finally, the vertices of the form $ZWH$ can be read from the Lagrangian
\begin{equation}
{\cal L} = -{g_R^2\over \sqrt{2}}{\sin\theta_R  \over 
\sin\theta_W}{v v_R \over \sqrt{v^2 + v_R^2}}  Z^\mu W_{R\mu}^\pm H^\mp_2
\end{equation}

\subsection{Loops involving Gauge Bosons}

Here we present results for the individual diagrams in Figure~\ref{diagsw}. 

\subsection*{Diagram \ref{diagsw}~a}

To check our results we first evaluate this diagram for the case 
of the Standard Model in Unitary Gauge. In this case the internal 
wavy lines are $W^+$ and $W^-$ and one finds,
\begin{eqnarray}
\Gamma_\mu &=& -i {g \over 2 \cos\theta_W} \gamma_\mu 
{g^2 \cos^2\theta_W \over 16 \pi^2} \nonumber \\
&&
\left({1\over \epsilon} \left({-M_Z^4 \over 12 M_W^4} -{4 M_Z^2 \over 3 M_W^2}
+{M_t^2 M_Z^2 \over 4 M_W^4} + {3 M_t^2 \over 2 M_W^2}\right) 
+fa_L \right) P_L
\label{l3g}
\end{eqnarray}
Notice that the first two divergent terms in 
Eq.~\ref{l3g} that are not proportional to $M_t^2$ are not included in 
the finite quantity $\delta g_{Lb}$ of Eq.\ref{smf}. These divergent 
terms cancel against other contributions from the renormalization of 
$G_F$, $M_Z$ and $\alpha$ when one calculates observables such as the 
partial $Z$ width \cite{Dawson:1994fa}. 
It is possible to obtain a simple expression for the finite part $fa_L$
in the limit  $M_W^2/M_t^2 <<1$. It is given by,
\begin{equation}
fa_L \approx -{M_t^2\over 8 M_W^2} \left(
{1\over \cos^2\theta_W}\left(2\log\left({M_t^2\over M_Z^2}\right)-3\right)
+12\log\left({M_t^2\over M_Z^2}\right) -10\right)
\end{equation}

For our model, the terms that are enhanced by $\cot^2\theta_R$ are 
obtained when the internal wavy lines are $W_R^+$ and $W_R^-$, resulting in,
\begin{eqnarray}
\Gamma_\mu &=& -i {g \over 2 \cos\theta_W} \gamma_\mu 
{-g_R^2 \sin^2\theta_W \over 16 \pi^2} \nonumber \\
&&
\left({1\over \epsilon} \left(
{-M_Z^4 \over 12 M_{W_R}^4} -{4 M_Z^2 \over 3 M_{W_R}^2}+
{M_t^2 M_Z^2 \over 4 M_{W_R}^4} + {3 M_t^2 \over 2 M_{W_R}^2}\right) 
+fa_R \right) P_R
\label{r3g}
\end{eqnarray}
The first two divergent terms in this expression, the ones not 
proportional to $M_t^2$, do not contribute to the vertex correction 
$\delta g_{Rb}$ that we are computing because they are not enhanced 
by $\cot^2\theta_R$. Although they appear to be proportional to $g_R^2$, 
they really are not when one considers the relations from Eq.~\ref{simpmass}. 
The finite part can be calculated numerically (we only present these 
results for the sum of all diagrams). For example, for $M_{W_R}=500$~GeV, 
we find $fa_R = 1.32$ with a renormalization scale $\mu=M_Z$.

It is possible to present approximate analytical results by taking 
$M_Z =0$ in the integrals. Doing so and expanding the resulting 
expression in powers of $M_t^2/M_{W_R}^2$ we find,
\begin{equation}
-i {g_L \over 2 \cos\theta_W}{g_R^2 \over 16 \pi^2}(-\sin^2\theta_W)
\left[{3\over 2}{M_t^2\over M_{W_R}^2}{1\over \epsilon}+{3\over 2}+
{M_t^2\over M_{W_R}^2}\left({11\over 4}-{3\over 2}\log\left(
{ M_{W_R}^2\over \mu^2}\right)\right)\right]
\end{equation}

\subsection*{Diagram \ref{diagsw}~b}

Once again we first evaluate Diagram b in Figure~\ref{diagsw} for the 
Standard Model in unitary gauge. In that case the internal wavy line 
is a charged $W$ and the intermediate state quarks are top. The result 
can be written in the limit  $M_W^2/M_t^2 <<1$ as,
\begin{eqnarray}
\Gamma_\mu &=& -i {g \over 2 \cos\theta_W} \gamma_\mu 
{g^2 |V_{tb}|^2\over 16 \pi^2} {M_t^2\over M_W^2} 
\left( 
\left(-{1\over \epsilon} \log\left({M_t^2\over M_Z^2}\right) -{1\over 2}
 \right) g_{Lt} \right. \nonumber \\
&+& \left.
\left({1\over 4 \epsilon}-{1\over 4}\log\left({M_t^2\over M_Z^2}\right)
-{1\over 8}
 \right) g_{Rt} \right) P_L\, .
\label{ltb}
\end{eqnarray}

For our model the only terms enhanced by $\cot^2\theta_R$ 
are obtained with the intermediate wavy line representing a $W_R^\pm$ and 
the intermediate state quark being top. As before, we present an 
approximate analytical result obtained by setting $M_Z=0$ in the loop 
integrals and expanding the resulting expression in powers of
$M_t^2/M_{W_R}^2$. This yields,
\begin{eqnarray}
&&-i {g_L \over 2 \cos\theta_W}{g_R^2 \over 16 \pi^2}\left\{
{1\over \epsilon}{M_t^2\over M_{W_R}^2}
\left({g_{Lt} \over 4}-g_{Rt}\right) 
-{3\over 4} g_{Rt} \right.\nonumber \\
&&\left. +{M_t^2\over M_{W_R}^2}\left[
\left(\log\left({ M_t^2\over \mu^2}\right)-{1\over 2}\right)g_{Rt}-
{1\over 4}\left(\log\left({ M_t^2\over \mu^2}\right)
+3\log\left({ M_t^2\over M^2_{W_R}}\right)+{7\over 2}\right)g_{Lt}
\right]\right\}
\end{eqnarray}
We only consider the case with $M_Z\neq 0$ numerically and include 
it in the sum of all diagrams in Figure~\ref{numres}.

\subsection*{Diagram \ref{diagsw}~c}

Finally we evaluate the wave-function renormalization diagrams of 
Figure~\ref{diagsw}~c. Once again we begin by considering the Standard 
Model in unitary gauge. For this diagram we can present an exact 
analytical result,
\begin{eqnarray}
1-Z_b^{-1} &=& {g^2 \over 16 \pi^2} P_L
\left( {3 M_t^2 \over 4 M_W^2} {1\over \epsilon}  -
{3 M_t^2 \over 4 M_W^2}\log\left({M_t^2\over \mu^2}\right) 
\right. \nonumber \\
&-& \left. 
{3 M_t^2 M_W^2 \over 4(M_t^2-M_W^2)^2}\log\left({M_W^2\over M_t^2}\right)
-{6M_W^4+5M_t^2M_W^2-5M_t^4 \over 8 M_W^2(M_t^2-M_W^2)}\right)
\label{wfr}
\end{eqnarray} 

Similarly, for the case of our model, the terms that are enhanced by 
$\cot^2\theta_R$ are obtained from the exchange of charged $W_R^\pm$ 
gauge bosons and we find,
\begin{eqnarray}
1-Z_b^{-1} &=& {g_R^2 \over 16 \pi^2} P_R
\left( {3 M_t^2 \over 4 M_W^2} {1\over \epsilon}  -
{3 M_t^2 \over 4 M_W^2}\log\left({M_t^2\over \mu^2}\right) \right. \nonumber \\
&-& \left. 
{3 M_t^2 M_W^2 \over 4(M_t^2-M_W^2)^2}\log\left({M_W^2\over M_t^2}\right)
-{6M_W^4+5M_t^2M_W^2-5M_t^4 \over 8 M_W^2(M_t^2-M_W^2)}\right)
\label{wfrr}
\end{eqnarray} 

As a check of our calculation we have evaluated the corresponding 
expressions for the Standard Model in unitary gauge. From these we 
can obtain, by adding the three contributions, the vertex correction 
terms $\delta g_{Lb}$ proportional to $M_t^2$. This result, Eq.~\ref{smf}, 
is finite and in agreement 
with the known result \cite{Altarelli:hv}. 

\subsection{Loops with Scalar Mesons}

We now turn our attention to the diagrams in Figure~\ref{hdiags}. 
As described in the main text we use the basis of Eq.~\ref{scalarbasis} 
assuming all scalars to be degenerate and to have a large mass.

\subsection*{Diagram \ref{hdiags}~a}

This type of diagram involves one gauge boson and one physical charged 
scalar in the loop. There is only a contribution to the right-handed 
coupling involving the $W_R^\pm H_2^\mp$ intermediate state. Our result 
in the limit where the Higgs masses are much larger than other masses is:
\begin{equation}
-i{g_L \over 2 \cos\theta_W}{1\over 16 \pi^2}\left({m_t\over v}\right)^2
\left({1\over \epsilon}+{3\over 2}-\log\left({M_H^2\over
      \mu^2}\right)\right)x_v(1-x_v)P_R
\end{equation}

\subsection*{Diagram \ref{hdiags}~b}

This type of diagram involves two scalars in the intermediate state. The 
left-handed coupling receives contributions from charged $H_1$ scalars as 
well as from a diagram with one neutral $H_1^0$ pseudo-scalar and one 
$h_2$ neutral scalar. The right-handed coupling receives contributions 
from the same diagram with neutral scalars as well as from the diagram 
with charged $H_2$ scalars.
\begin{eqnarray}
&&-i{g_L \over 2 \cos\theta_W}{1\over 16 \pi^2}\left({m_t\over v}\right)^2
\left({1\over \epsilon}+{1\over 2}-\log\left({M_H^2\over
      \mu^2}\right)\right)
\nonumber \\
&&\left[ \left(\sin^2\theta_W + \cos^2\theta_W x_v 
-{1\over 2} x_v^2 \right) P_R +
\left( -{1\over 2}-{\cos^2\theta_R\over 2}\cos 2\theta_W
\right)P_L\right]
\end{eqnarray}

\subsection*{Diagram \ref{hdiags}~c}

This type of diagram involves the exchange of a charged $H_1$ or of a 
neutral $H_1^0$ or $H_2$ scalars for the left-handed coupling as well 
as exchanges of a charged $H_2$ scalar or a neutral $A_1^0$ or $H_2$ scalars 
for the right-handed coupling.
\begin{eqnarray}
-i{g_L \over 2 \cos\theta_W}{1\over 16 \pi^2}\left({m_t\over v}\right)^2
\left({1\over \epsilon}+{1\over 2}-\log\left({M_H^2\over
      \mu^2}\right)\right)
\nonumber \\
\left[ \left(-{1\over 3}\sin^2\theta_W + (-{1\over 2}+{2\over
      3}\sin^2\theta_W)x_v \right) P_R +
\left({1\over 3}\sin^2\theta_W (1-2\cos^2\theta_R)\right)P_L 
\right]
\end{eqnarray}

\subsection*{Diagram \ref{hdiags}~d}

This diagram represents $b$ wave-function renormalization through 
scalar loops. Once again, for the left-handed coupling one obtains 
contributions from exchanging a charged $H_1$ and neutral $H_1^0$ 
and $H_2$, whereas for the right-handed coupling the contributions 
arise through exchange of charged $H_2$ and neutral $H_1^0$ and $H_2$
scalars.
\begin{eqnarray}
-i{g_L \over 2 \cos\theta_W}{1\over 16 \pi^2}\left({m_t\over v}\right)^2
\left({1\over \epsilon}+{1\over 2}-\log\left({M_H^2\over
      \mu^2}\right)\right)
\nonumber \\
\left[ \left(-{2\over 3}\sin^2\theta_W +{1\over 3}\sin^2\theta_W x_v \right) P_R +
\left((1+\cos^2\theta_R)({1\over 2}-{1\over 3}\sin^2\theta_W\right) 
P_L\right]
\end{eqnarray}

\end{document}